\documentclass[aps,pra,twocolumn,superscriptaddress,amsmath,amssymb,floatfix]{revtex4}

\usepackage{dcolumn}
\usepackage{bm}
\usepackage{dsfont}
\usepackage{amsbsy}
\usepackage{graphicx,times}
\usepackage{appendix}

\date{\today}

\begin{document}

\title{Proposal for enhanced resolution in nanoscale NMR:  quantum sensing with pulses of finite duration}

\author{J. E. Lang}
\affiliation{Department of Physics and Astronomy, University College London, Gower Street, London WC1E 6BT, United Kingdom}

\author{J. Casanova}
\affiliation{Institut f\"{u}r Theoretische Physik and IQST, Albert-Einstein-Allee 11, Universit{\"a}t Ulm, D-89069 Ulm, Germany}

\author{Z.-Y. Wang}
\affiliation{Institut f\"{u}r Theoretische Physik and IQST, Albert-Einstein-Allee 11, Universit{\"a}t Ulm, D-89069 Ulm, Germany}

\author{M. B. Plenio}
\affiliation{Institut f\"{u}r Theoretische Physik and IQST, Albert-Einstein-Allee 11, Universit{\"a}t Ulm, D-89069 Ulm, Germany}

\author{T. S. Monteiro}
\affiliation{Department of Physics and Astronomy, University College London, Gower Street, London WC1E 6BT, United Kingdom}

\begin{abstract}

The nitrogen vacancy (NV) color center in diamond is an enormously important platform for the development of quantum sensors, including for single spin and single molecule NMR. Detection of weak single-spin signals is greatly enhanced by repeated sequences of microwave pulses; in these dynamical decoupling (DD) techniques, the key control parameters swept in the experiment are the time intervals, $\tau$, between pulses. Here we show that, in fact, the pulse {\em duration} $t_p$ offers a powerful additional control parameter. While previously, non-negligible $t_p$ has been considered simply a source of experimental error, here we elucidate the underlying quantum dynamics: we identify a landscape of quantum-state crossings which are usually closed (inactive) but may be controllably activated (opened) by adjusting $t_p$ from zero. We identify these crossings with recently observed but unexpected dips (so called “spurious dips”) seen in the quantum coherence of the NV spin. With this new understanding, both the position and strength of these sharp features may be accurately controlled; they co-exist with the usual broader coherence dips of short-duration microwave pulses, but their sharpness allows for higher resolution spectroscopy with quantum diamond sensors, or their analogues.
 
\end{abstract}

\maketitle

\section{Introduction}

The nitrogen-vacancy (NV) colour center in diamond is a powerful nanoscale probe of its local environment \cite{schirhagl2014nitrogen, rondin2014magnetometry, wu2016diamond}. The spin-1 electronic ground state can be initialised and measured using lasers via optically detected magnetic resonance providing a window into its surroundings. A wide range of sensing techniques have been developed to perform the detection of individual nuclear spins \cite{zhao2012sensing,taminiau2012detection,kolkowitz2012sensing, sushkov2014magnetic, muller2014nuclear, mkhitaryan2015highly}, spin clusters \cite{zhao2011atomic,shi2014sensing, ma2016angstrom, wang2016positioning} and even their manipulation for use as quantum registers \cite{cappellaro2009coherence, neumann2010quantum, taminiau2014universal, casanova2016noise, liu2013noise}.

For nuclear spin detection, a sequence of microwave $\pi$ pulses repeatedly inverts the electronic state of the NV. If the pulse rate exceeds the width of the environmental spectral density this permits effective decoupling from magnetic noise arising from the surrounding bath of nuclear spins, extending electronic coherence times by orders of magnitude \cite{de2010universal, du2009preserving}. However, wherever the pulse spacing $\tau$ is resonant with the target nuclear spins, the converse is true: system-environment interactions are magnified and a sharp signal is observed in NV coherence measurements.
The above mechanism underpins a wide range of techniques and pulse protocols 
termed dynamical decoupling (DD).  

The effectiveness of DD has stimulated interest in new types of
 pulse sequences to obtain increased resolution of experimental signals and protect against pulse errors \cite{casanova2015robust, zhao2014dynamical, ma2015resolving}. The free parameters considered in these experiments have so far been only the pulse spacings and pulse phases. However, application of certain pulse sequences has been shown to produce so called `spurious' signals due to the finite duration of pulses \cite{loretz2015spurious, haase2016pulse}. To date these have been considered sources of error as they lead to ambiguities in nuclear spin classification presenting a considerable problem for nanoscale NMR and MRI experiments.

Experimental analysis of DD has commonly been treated with semi-classical noise theory \cite{cywinski2008enhance, zhao2012sensing} and geometric approaches \cite{taminiau2012detection, kolkowitz2012sensing} which are exact only in the simplest cases. Floquet analysis of the full quantum dynamics was recently introduced  \cite{lang2015dynamical} as a natural framework of analysis. The method related an observed coherence dip to avoided crossings
in the underlying quantum eigenstates: the shape and depth of coherence dips are determined by the width of each avoided crossing.

In the present work we show that the avoided crossings which produce
the usual coherence dips of quantum spin sensing are actually part of a much 
larger family
of crossings. For negligible pulse duration ($t_p=0$), these are true crossings
and thus yield no signal. We show here that the effect
of $t_p>0$ is to open, and thus to activate (to a varying extent), this larger
landscape of crossings. We develop an accurate quantum model which 
enables us to calculate reliably the strength of these new signals.
We identify these opened crossings with the `spurious' dips  identified in recent experiments \cite{loretz2015spurious}.

 By understanding the underlying quantum dynamics we are able to accurately model these newly identified signals: for small $t_p$, anti-crossings
 are narrow so the experimental signals are sharply peaked. We can then propose
 how to exploit the pulse width $t_p$ as a new experimental parameter
which offers an increase in resolution without the need for complex pulse sequences, i.e. utilising the commonplace XY family \cite{gullion1990new} of pulse sequences. Furthermore, we show that spurious signals can either be enhanced or suppressed, i.e. controlled, allowing the unambiguous classification of nuclear spins. This work is not only limited to experiments with NV centers as the theory is general, it could be applied for other defect centers, such as those in silicon carbides.

\begin{figure*}
\begin{center}
\includegraphics[width=178mm]{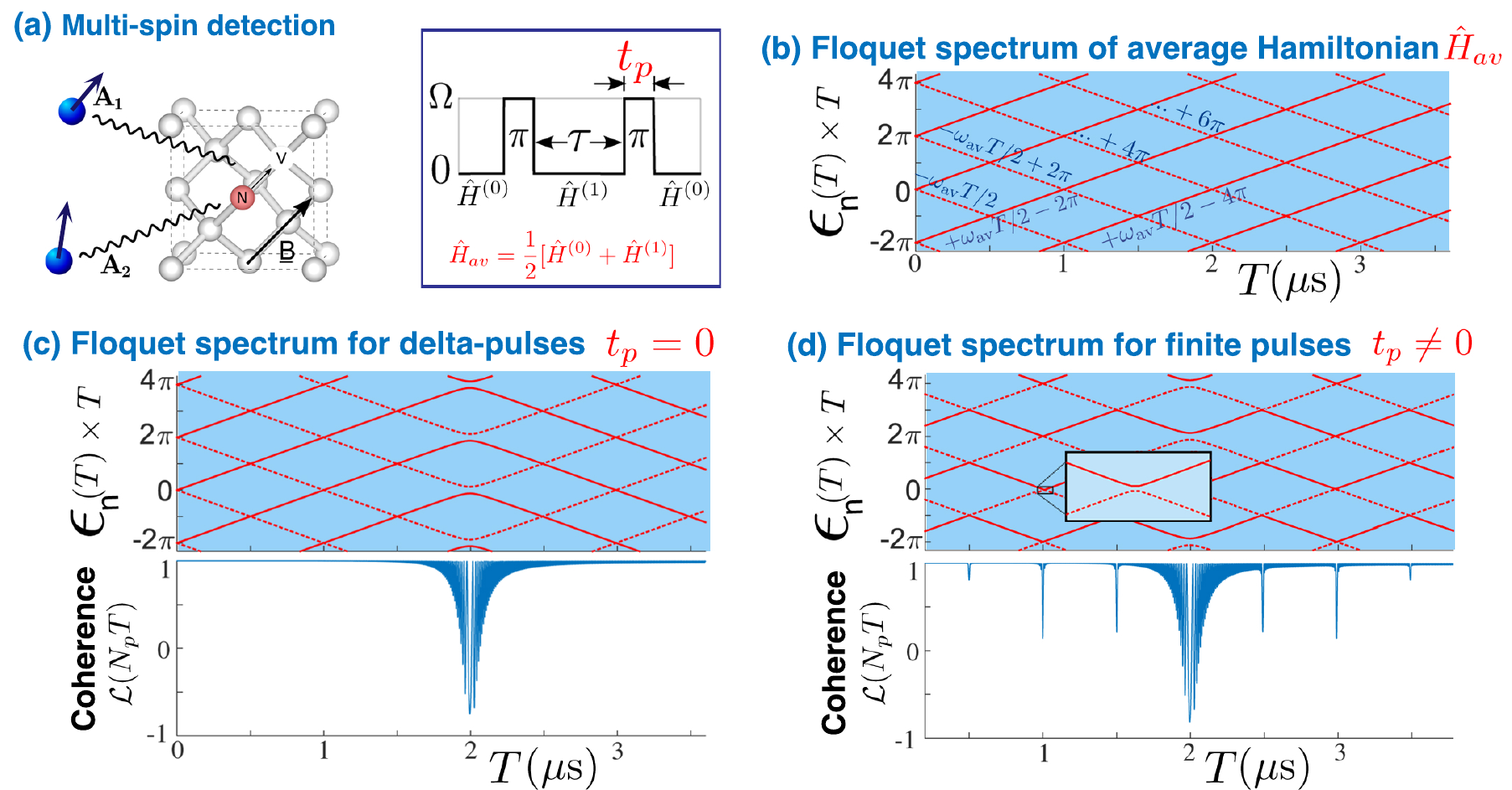}
\caption{{\bf (a)} Dynamical decoupling (DD) pulses sequences are commonly applied in sensing protocols using NV centers. A series of microwave pulses repeatedly flip the electronic spins between an  
$  |m_s=\pm 1\rangle$  and a $ |m_s=0\rangle$ state. A frequent convenient approximation is to assume that the dynamics of an environmental spin is given from a period-averaged Hamiltonian
$\hat{H}_\text{av}$ (see Section~\ref{sec: FloqHam} for details) with dips in coherence occurring at pulse spacing $\tau_{dip} \sim \pi/\omega_\text{av}$. We demonstrate in Section \ref{sec: Resolution} that spurious dips can be used to increase resolution for the detection of multiple isolated spins. ($\hat{H}^{(m_s = 0,1)}$ is the (NV state dependent) nuclear Hamiltonian.)
{\bf (b)} The  Floquet eigenspectrum for $\hat{H}_\text{av}$ is given by $\omega_\text{av}$, but shows the characteristic Floquet structure of dressed states shifted by integer multiples of $\omega=2\pi/T $ (sometimes termed `multi-photon states' ). In this case, level degeneracies correspond to true crossings.
{\bf (c)} The effect of the DD pulses, for the ideal $t_p = 0$ case, is to turn some crossings into avoided crossings. A coherence dip, of strength determined by the width of the crossing, can then be seen.
{\bf (d)} For the general $t_p\neq 0$ case, all remaining crossings can potentially become avoided crossings: these are typically narrow, and yield sharp weaker dips. Our main finding is to show that we
can determine these analytically; a key result of this work  is Eq.~\eqref{eq: Lspurious} where we give a closed-form expression for the shape and strength of these so-called `spurious' coherence dips and argue they are in fact useful for sensing. For numerical simulations here: $\omega_\text{av} = 2\pi\times  2\text{ MHz}$ and $A_\perp = 2\pi\times 200\text{ kHz}$. The pulse sequence is XY8 with $N_p = N/8 = 60$ repetitions. The finite pulses have height $\Omega =2\pi\times   20\text{ MHz}$ and duration $t_p = \pi/\Omega$. $T$ is in microseconds.}
\label{Fig1}
\end{center}
\end{figure*}

The physical systems of interest here are modelled by 
 temporally periodic Hamiltonians, $\hat{H}(t) = \hat{H}(t + T)$, where $T$ is the period. For example, for an XY4 NMR microwave pulse sequence, $T=4\tau$, which is then repeated $N_p$ times. It can be shown that the behaviour is given by an eigenvalue equation, 
\begin{equation}
\hat{H}_F|\Phi^{(n)}_F\rangle = \epsilon_{n} |\Phi^{(n)}_F\rangle,
\label{Floqueteq}
\end{equation}
quite analogous to a time-independent (but infinite) Schr\"{o}dinger equation, but where 
in place of a Hamiltonian, we diagonalise instead the (Hermitian) Floquet operator:
\begin{equation}
\hat{H}_F= \left[\hat{H}(t) - i \frac{\partial}{\partial t}\right]
\label{FloquetH}
\end{equation}
to obtain the corresponding Floquet states $|\Phi^{(n)}_F\rangle$ and Floquet quasienergies $\epsilon_{n}$.

Previously \cite{lang2015dynamical}, the simple case of  nuclear spin detection with a Carr-Purcell-Meiboom-Gill (CPMG) sequence \cite{carr1954effects, meiboom1958modified} of idealised, infinitely-sharp $\delta$-pulses (where the pulse duration $t_p = 0$), allowed certain simplifications: it was in fact straightforward to  obtain Floquet states and quasienergies by direct diagonalisation of the one-period unitary  $\hat{U}(T)$, constructed by concatenating the evolution between pulses.
For general pulses, this procedure would not be adequate; e.g. for pulses of finite duration, the eigenvectors and quasienergies $\epsilon_{n} \equiv \epsilon_{n}(t_p)$ now depend  on pulse duration; construction of the one-period evolution by concatenation becomes cumbersome. In the present work, it becomes essential to obtain a
full diagonalisation of the Floquet operator with Eq.~\eqref{FloquetH}.

When solving Eq.~\eqref{Floqueteq}, a usual procedure is to write the Hamiltonian:
\begin{equation}
\hat{H}(t)= \hat{H}_0 + \hat{H}_p(t), 
\label{CanonH}
\end{equation}
in terms of a time-independent component $\hat{H}_0$ and a time-periodic potential
$ \hat{H}_p(t)=\hat{H}_p(t+T)$. For nanoscale NMR applications, $\hat{H}_0$ would represent the spin Hamiltonians, including Zeeman terms and spin-spin interactions; while 
$ \hat{H}_p(t)$ would represent the effect of the microwave pulses. Typically, the Floquet operator is diagonalised in the basis of $\hat{H}_0$. While this allows accurate numerical solutions, it offers comparatively little physical insight.\\

However, the key part of our approach is to show that we can consider instead an equivalent 
rotating-frame Hamiltonian:
\begin{equation}
\hat{H}'(t)= \hat{H}_\text{av} + \hat{V}_{t_p}(t), 
\label{RotFH}
\end{equation}
where $\hat{H}_\text{av}$ is time-independent, \emph{average Hamiltonian}, while $\hat{V}_{t_p}$ is an
effective time-periodic potential, dependent on pulse duration, that couples the average Hamiltonian eigenstates. Section~\ref{sec: FloqHam} details the transformation to this frame.

In Fig.~\ref{Fig1}, we illustrate the Floquet spectrum and the associated coherence behaviour.
The average spin Hamiltonian in its eigenbasis is 
$\hat{H}_\text{av} = \sum_{\alpha} |\alpha\rangle\langle\alpha| \omega_\text{av}^\alpha$.

{\bf (i)} For the  case $ \hat{V}_{t_p} = 0$, the unperturbed Floquet spectrum is shown in Fig.~\ref{Fig1}(b) and its eigenvalues are  given by $\omega_\text{av}^\alpha + l\omega$ where $\alpha = 1,...,D$, with $D$ the dimension of the spin system. The additional index $l= 0,\pm 1,\pm 2,...$ is a consequence of the invariance of Floquet  states to translations in $\epsilon_n$ by integer multiples of $\omega = 2\pi/T$. The Floquet theorem is the temporal analogue of the Bloch theorem and analogously to the band structure seen in lattice dynamics in condensed matter systems, the Floquet eigenspectrum has additional levels structures,
corresponding to quasienergy shifts of single quanta in $\omega$. Hence the unperturbed spectrum has degeneracies wherever a pair of eigenvalues $\omega_\alpha , \omega_{\alpha'}$ differ by an integer multiple of $\omega$ (see below).  However, in this case, degeneracies yield true crossings and no coherence dip is seen.

{\bf (ii)}  For the case $\hat{V}_{(t_p= 0)}$, when we apply $\pi$-pulses
which approximate ideal $\delta$-spikes, some crossings become anti-crossings and the usual structure
of coherence dips appears. The width
of the crossing (and hence dip shape and visibility) depends crucially on  the matrix elements of the Floquet operator. In the dressed state basis where $|l\rangle = e^{il\omega t}$:
\begin{multline}
\label{eq: FloquetHam}
\langle m_s \alpha l|\hat{H}_F|m_s' \alpha' l'\rangle =  
(\omega_\text{av}^\alpha + l\omega)\delta_{\alpha \alpha'}\delta_{ll'}\delta_{m_s m_s'}  \\
+  \langle m_s \alpha l |  \hat{V}_{t_p} |m_s'\alpha' l'\rangle.
\end{multline}   
Whether or not a dip is seen can be understood, to first order, if we consider the
diagonal contributions to be dominant so the levels may still be labelled by the unperturbed
basis $m_s \alpha l$. If the off-diagonal matrix element of  $\hat{V}_{t_p} $ is non-zero, then the degeneracy between  the corresponding levels will be lifted, creating an avoided crossing which causes a coherence dip. For  $t_p=0$  only $\langle m_s \alpha l | \hat{V}_{t_p}  |m_s' \alpha' l'\rangle \delta_{m_s m_s'}\neq 0 $: for the usual dynamical decoupling scenario, the bath evolution in the two subspaces corresponding to the electronic states of the NV sensor remains fully independent.

{\bf (iii)} Finally, we consider and illustrate in Fig.~\ref{Fig1}(c) the general case $t_p \neq 0$ where we apply $\pi$-pulses of finite duration. In this case we find there are non-zero couplings for the case $m_s \neq m_s'$ so here potentially, all crossings are avoided crossings. These connect  previously uncoupled 
(albeit degenerate) levels, hence turning true crossings into avoided crossings and generating new, `spurious', coherence dips. 

  The paper is constructed as follows: In the next section (Section \ref{sec: FloqHam}) we explain how to construct the Floquet Hamiltonian matrix and in particular, how to calculate
explicitly the matrix elements of $\hat{V}_{t_p}$. We exploit the fact that the dynamics at an avoided crossing  reduces to an effective two-state system in order to obtain analytical forms for the coherence functions. In Section \ref{sec: SingleSpins} we apply the method to the sensing of hyperfine coupled spins  and show how to
obtain analytically the shape and magnitude of the coherence dip for pulses of arbitrary duration $t_p$ and for pulse sequences with arbitrary pulse positions. Tuning the pulse sequence can open and close avoided crossings thus allowing new control over both expected  and spurious coherence dips, and hence the effective coupling to nuclear targets. A key result of this work is an expression for spurious coherence dips  given in Eq.~\eqref{eq: Lspurious} and shown, in Section \ref{sec: Numerics}, to be in excellent agreement with full numerics. In Section \ref{sec: Resolution} we demonstrate how the small width of spurious avoided crossings equates to a significant increase in spectral resolution and in Section \ref{sec: Conclusion}, we conclude. 

\section{Constructing the Floquet Hamiltonian}
\label{sec: FloqHam}
\begin{figure}[ht!]
\begin{center}
\includegraphics[width = 82mm]{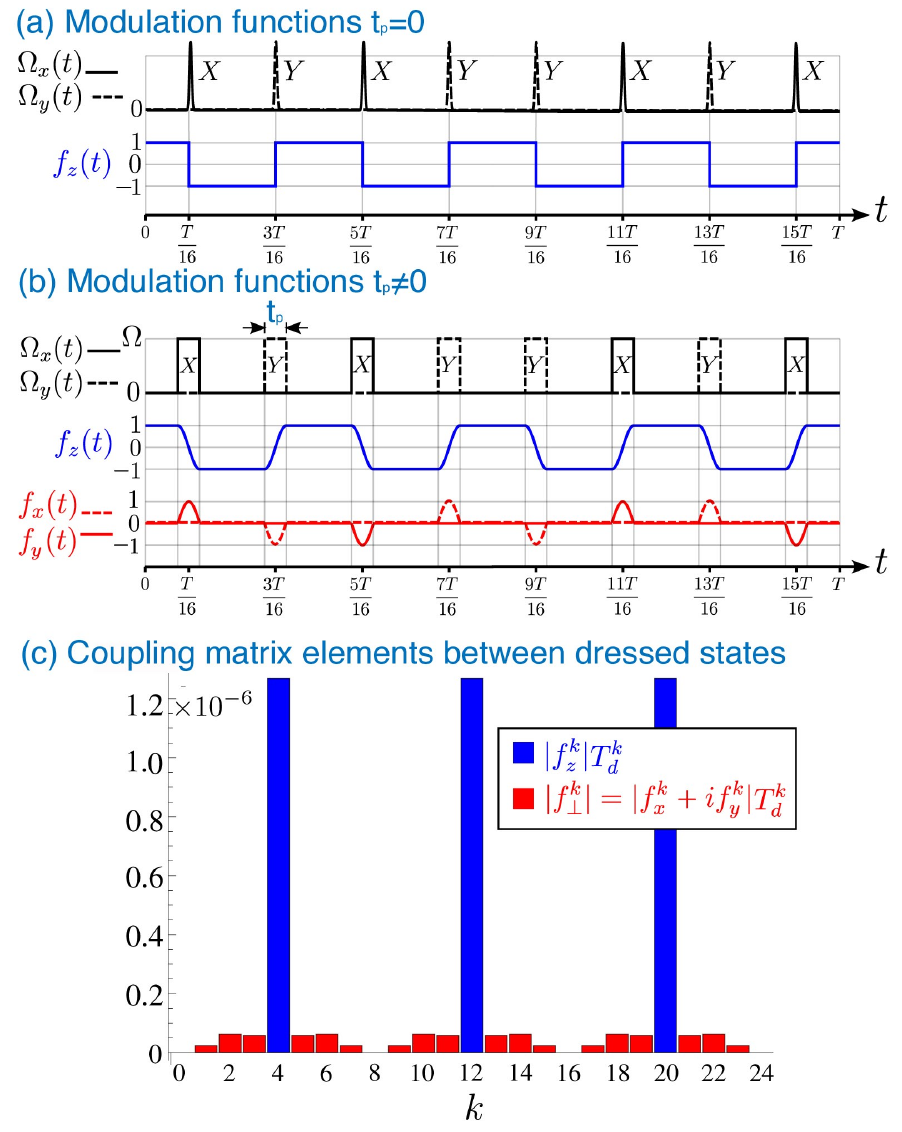}
\caption{ Shows the modulation functions for an XY8 sequence with {\bf (a)} ideal $\pi$ pulses and {\bf (b)} pulses  
which provide a $\pi$ rotation, but are of  finite duration $t_p$. For this case, there are non-zero 
contributions from $f_{x,y}(t)$ modulation functions which can couple the $m_s$ subspaces.   
In our quantum Floquet treatment, the modulation functions replace and generalise the role played by 
the filter function in classical noise treatments of dynamical decoupling-based sensing.
{\bf (c)} Matrix elements of the modulations function $\langle l+k|f_{i}|l\rangle$ in the dressed state basis. The $f_{z}$ matrix elements (blue) for $k=4,12,20...$ are responsible for the `expected' coherence dips and are of much larger magnitude than the $f_{x,y}$ matrix elements which are only non-zero for $t_p>0$
and give rise to the so-called `spurious' coherence dips. CPMG control does not produce spurious signals precisely because in that case the non-zero $f_\perp^k$ and $f_z^k$ coincide. This is discussed in Appendix~\ref{app: CPMG}.} 
\label{fig: pulses}
\end{center}
\end{figure}

For the detection of nuclear spin clusters via an NV center, it is usual to give the Hamiltonian
in the form:
\begin{equation}
\hat{H}(t) = \hat{H}_0 + \hat{H}_p(t)\equiv
\hat{\mathds{I}} \otimes \hat{H}_\text{av} + \hat{\sigma}_z \otimes \hat{V} + \hat{H}_p(t),
\end{equation}
which is in the form of Eq.~\eqref{CanonH}. A distinction is now made between the NV and nuclear spin subspaces. Here,   $\hat{\mathds{I}}$ and $\hat{\sigma}_z$ are Pauli matrices in the NV $m_s = 0,1$ subspace, $\hat{H}_\text{av}\equiv\frac{1}{2}[\hat{H}^{(0)} + \hat{H}^{(1)}]$ and 
$\hat{V}\equiv\frac{1}{2}[\hat{H}^{(1)}- \hat{H}^{(0)}]$ are the nuclear average and interaction Hamiltonians respectively. Here $\hat{H}^{(m_s = 0,1)} = \langle m_s|\hat{H}_0|m_s \rangle$ is the nuclear spin Hamiltonian conditioned on the NV state, see Appendix~\ref{app: AvHam} for more details.

In the frame rotating at the microwave frequency the pulse sequence Hamiltonian takes the form:
\begin{equation}
\hat{H}_p(t) =  \Omega_x(t)\hat{S}_x + \Omega_y(t)\hat{S}_y,
\label{PulseH}
\end{equation}
where $\hat{S}_{x,y}$ are the NV spin operators and $\Omega_{x,y}(t)$ describes the amplitude of the microwave driving as a function of time. 

In order to obtain the Hamiltonian in the required form of Eq.~\eqref{RotFH}, we introduce a frame that rotates under $\hat{H}_p(t)$. That is  $ \hat{H}'(t) = \hat{U}^\dagger_p(t)\hat{H}_0\hat{U}_p(t)$ where $\hat{U}_p(t) = \mathcal{T}\exp[-i \int_0^t \hat{H}_p(t')dt']$ is the time-ordered propagator. 

The $\hat{H}_\text{av}$ component is unaffected by this transformation, but the remainder will yield the coupling matrix $\hat{V}_{t_p} $ of Eq.~\eqref{RotFH}:
\begin{equation}
 \hat{V}_{t_p}(t)=  \hat{U}^\dagger_p(t) \hat{\sigma}_z \hat{U}_p(t) \otimes \hat{V}  =  \sum_{i = x,y,z}f_i(t)\hat{\sigma}_i \otimes \hat{V}.
\label{Modfunct}
\end{equation}

Typically microwave pulses are modelled as infinitely sharp delta pulses. This approximation is adequate whenever $t_p$ is much less than the nuclear signal period, so that the detected spin states do not evolve appreciably over the duration of the pulse; and only a modest number of pulses sequence repetitions $N_p$ are applied so that  cumulative errors arising from these effects remain insignificant.

 In this case the coupling matrix reduces to $\hat{V}_{t_p}(t) = f_z(t)\sigma_z \otimes \hat{V}$, where $f_z(t)$ is the usual stepped modulation function (see Fig.~\ref{fig: pulses}(a)) defined in previous studies, and there are no $x$ or $y$ modulation functions. In this picture the new frame instantaneously flips along the $z$-axis at each pulse. We show here that the transformation to a rotating frame is equally applicable for pulses of finite width.

As found in recent studies \cite{loretz2015spurious} the finite duration of pulses must be taken into account, especially where increasing number of pulses are applied: the total phase accumulation time during pulses $Nt_p$ can become comparable or even greater than during $\omega_\text{av}^{-1}$ if the number of pulses, $N$, is sufficiently large. We now obtain the modulation functions for the finite-duration pulses.

\subsection{The $f_{x,y,z}(t)$ modulation functions for arbitrary $t_p$}

Although Eq.~\eqref{Modfunct} is generic, for convenience we test the method on  pulses  modelled by top-hat functions with some finite height $\Omega$ and width $t_p = \pi/\Omega$, chosen such that a full $\pi$-rotation is completed during the pulse. In this case during the pulse the frame does not instantaneously flip along the $z$-axis but must evolve smoothly between the initial and final positions. This induces a sinusoidal curve between $\pm 1$ during the pulse as seen in Fig.~\ref{fig: pulses}(b). The unitarity of the transformation to the rotating frame then requires that the transverse modulation functions, $f_{x,y}(t)$, be non-zero, again see Fig.~\ref{fig: pulses}(b). 

During the pulse the frame rotates about some axis in the $xy$-plane with a phase determined by the phase of the pulse being applied and also by the phases of the pulses that came before it. The rotation is such that a unit vector $+\hat{\textbf{z}}$ becomes $-\hat{\textbf{z}}$ by the end of the pulse.

For an isolated top hat pulse the transformation of $\hat{\sigma}_z$ is simple to calculate; however, the history of previous pulses must also be taken into account. Consider an arbitrary sequence of $n$ pulses, all of height $\Omega$ and length $t_p = \pi/\Omega$, applied at times $\{t_1,...,t_n\}$, and phases $\{\phi_1, ...,\phi_n\}$ ($\phi = 0$ and $\phi = \pi/2$ correspond to X and Y pulses respectively). During the $m$'th pulse, i.e. for $t' = t - t_m \in [-t_p/2,+ t_p/2]$
\begin{equation}
\hat{U}_p(t') = \exp(-i\Omega\hat{S}_{\phi_m} (t'+ tp/2))\times(-i\hat{\sigma}_{\phi_{m-1}})\times...\times(-i\hat{\sigma}_{\phi_{1}}).
\end{equation}
Thus we have 
\begin{multline}
\hat{U}^\dagger_p (t')\hat{\sigma}_z\hat{U}_p(t') = 
(-1)^m\sin\Omega t'\sigma_z   
+ \cos\Omega t' \cos\varphi_m\sigma_x   \\
+ \cos\Omega t' \sin\varphi_m\sigma_y,
\end{multline}
where $\varphi_m = 2\sum_{k=1}^{m-1}(-1)^{k+1}\phi_k + (-1)^{m+1}(\phi_m + \pi/2)$. The modulation functions during the $m$-th pulse are:
\begin{align}
\label{eq: modfuncs}
f_x(t') &= \cos\Omega t'\cos\varphi_m,   \nonumber \\
f_y(t') &= \cos\Omega t'\sin\varphi_m,   \nonumber \\
f_z(t') &= (-1)^m\sin\Omega t'
\end{align}
and between the pulses, $f_z(t) = \pm 1$ and $f_{x,y}(t) = 0$. 

In this section we have defined a new set of modulation functions for realistic pulse sequences that generalise the single step modulation function used to model infinitely sharp microwave pulses. These modulation functions can be calculated for any periodic pulse sequence and are crucial for deriving an analytical model of the spurious signals produced by realistic pulses.

\begin{figure}[ht!]
\begin{center}
\includegraphics[width = 80mm]{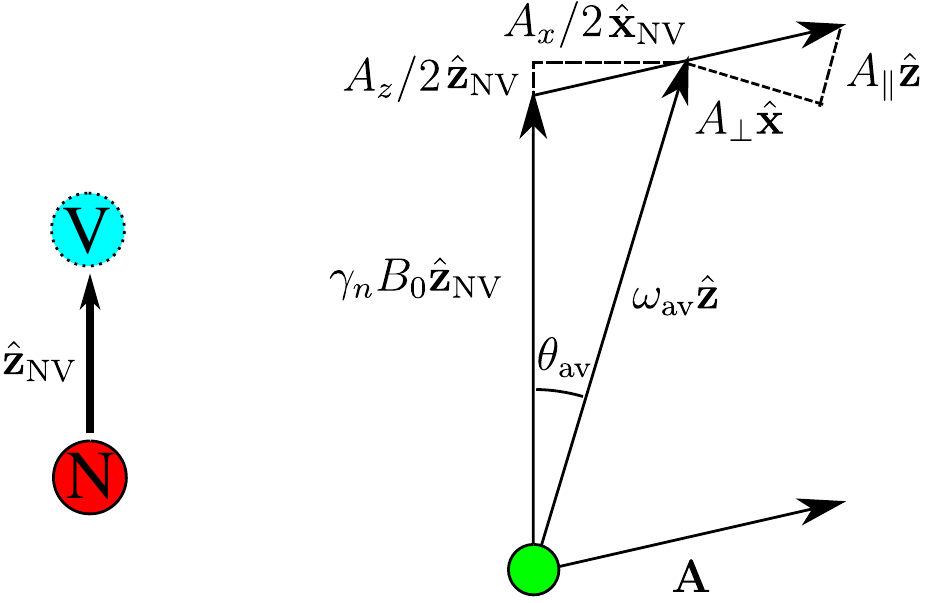}
\caption{The direction between the nitrogen and vacancy sites  in the NV center sets the axis $\hat{\textbf{z}}_\text{NV}$ and the external field, $\textbf{B} =  B_z\hat{\textbf{z}}_\text{NV}$, is applied parallel to this. The hyperfine field felt at the nucleus, $\textbf{A} = (A_x, 0, A_z)$, shifts the quantization axes of the nuclear Hamiltonian. It is useful to work in the nuclear average Hamiltonian basis and this quantisation axis is shown.}
\label{fig: physicalpic}
\end{center}
\end{figure}

\subsection{Matrix elements of the coupling $ \hat{V}_{t_p} $}

Now that the modulation functions $ f_i(t)$ have been obtained the matrix elements of $ \hat{V}_{t_p} $
may be straightforwardly evaluated  in the $|m_s \alpha l\rangle $ basis:

\begin{equation}
\langle m_s \alpha l | \hat{V}_{t_p}  |m_s' \alpha' l' \rangle = \sum_{i = x,y,z}f_i^{l-l'} \langle m_s | \hat{\sigma}_i |m_s' \rangle  \langle\alpha |\hat{V} |\alpha' \rangle.
\end{equation}
The $\langle m_s | \hat{\sigma}_i |m_s' \rangle$ are usual Pauli matrix elements while
evaluating  the nuclear interaction
Hamiltonian matrix elements $\langle\alpha |\hat{V} |\alpha' \rangle$  is a standard procedure
and typically quite straightforward:  for nuclear spin detection 
$\hat{V} = A_\perp\hat{I}_x$ (see \ref{app: coherence} for further details).

However,  the modulation function matrix elements in the dressed state basis are of most significance:
\begin{equation}
f_i^k = \langle l+k | f_i(t) | l \rangle = \frac{1}{T}\int_0^T dt' f_i(t')\exp(-ik\omega t'),
\end{equation}
 are the Fourier amplitudes of the periodic modulation functions. They play a key role in our approach
by selecting whether a crossing becomes an anti-crossing as well as determining its width; in effect they replace 
(and for the $f_\perp$ case generalise) the role
played by the filter function, a common method of analysis of spin sensing,
using classical noise models.

Importantly, for robust DD sequences (see Fig.~\ref{fig: pulses}), $f_z^k$ and $f_\perp^k = f_x^k + if_y^k$ are non-zero for different values of $k$. The $f_z^k$ terms create avoided crossings and coherence dips at `expected' pulse spacings whereas  $f_\perp^k$ creates new avoided crossings and dips at so-called `spurious' positions. The width of the avoided crossing is proportional to the magnitude of the relevant $f_z^k$ or $f_\perp^k$. The fact that $|f_z^k| \gg |f_\perp^k|$ immediately explains why the spurious dips are much sharper and typically only seen for large pulse numbers, it is a result of narrower avoided crossings. This new understanding of spurious dip sharpness leads us to propose a sensing protocol that exploits this behaviour to obtain increased spectral resolution. This is discussed in Section \ref{sec: Resolution}. The finite pulse effect in experiments using a CPMG sequence is discussed in Appendix~\ref{app: CPMG}.

\begin{figure*}[ht!]
\includegraphics[width=180mm]{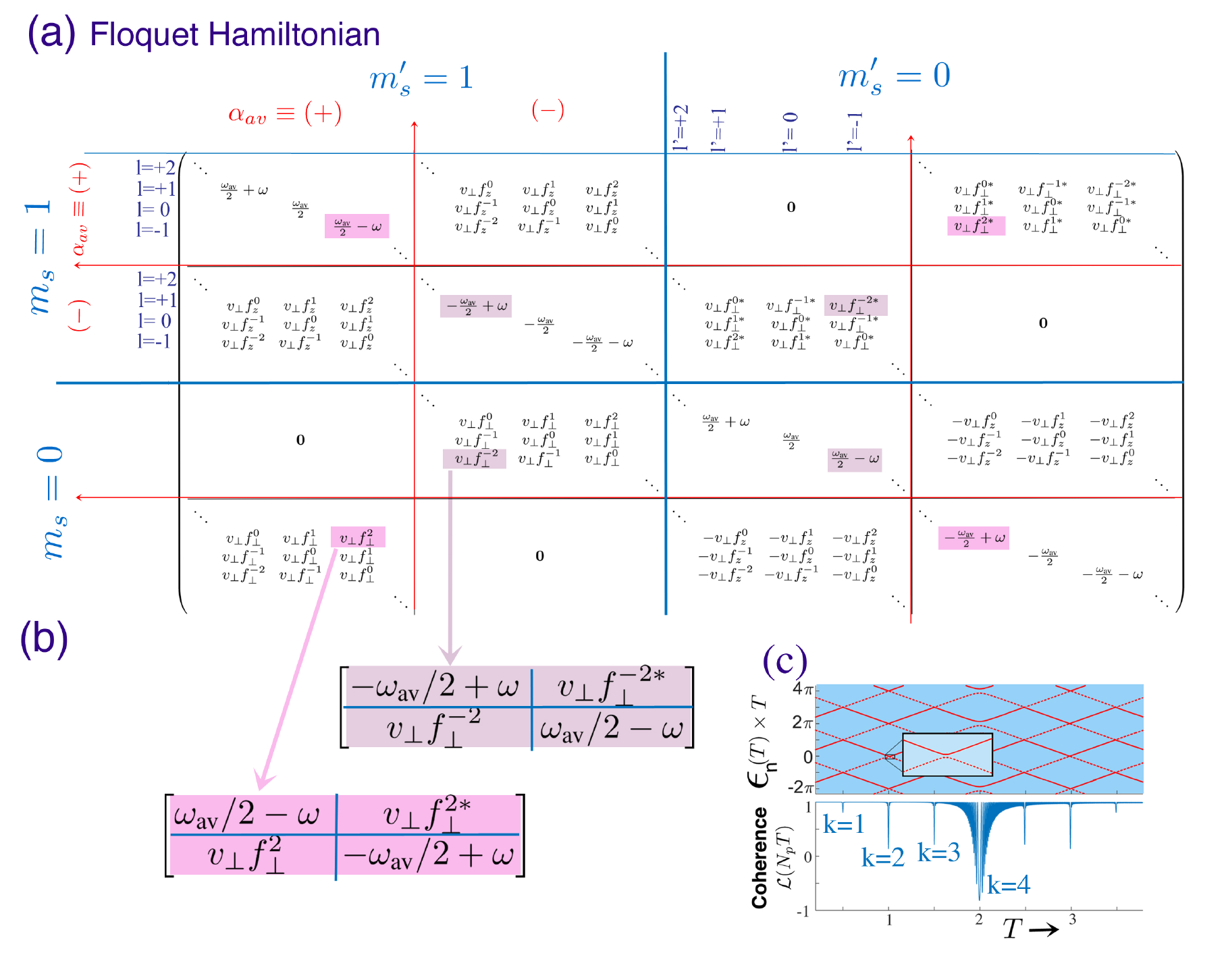}
\caption{Toolbox for calculating coherence function for finite pulses for detection of
 spin-half environmental spins. {\bf (a)} Schematic representation of the Floquet Hamiltonian matrix. Each element of the original Hamiltonian becomes an infinite block containing all the Fourier coefficients. The relative magnitude of the off-diagonal terms are determined by the hyperfine coupling strength $v_\perp = A_\perp/2$ and the Fourier series of the periodic modulation functions, $f_z^k$ (for $m_s=m_s'$ case, ordinary coherence dips) and $f_\perp^k = f_x^k + if_y^k$
(for $m_s\neq m_s'$, spurious dips). {\bf (b)} To evaluate the form of the $k=2$ dip, we consider the coupling, at $T \approx T_\text{dip}^{(k=2)}$, between the pairs of  states in the avoided crossing. The resulting $2\times 2$ matrices can be diagonalised exactly to yield the stroboscopic evolution (see main text Section \ref{sec: Coherence}) and an analytical expression for the shape of the coherence dip shown in {\bf (c)}.}
\label{fig:HamFloq}
\end{figure*}

\section{APPLICATION: single spin NMR}
\label{sec: SingleSpins}

\subsection{Average Hamiltonian states}

For the detection of a single spin 1/2, such as an environmental nuclear spin,  the average Hamiltonian  $\hat{H}_\text{av} = \omega_\text{av}\hat{I}_z$ and $\hat{V} = A_\perp\hat{I}_x$
hence the average Hamiltonian basis takes a very simple form.
 Fig.~\ref{fig: physicalpic} and Appendix~\ref{app: AvHam} describe how these relate to the Zeeman basis. In the weak coupling regime $A_\perp \ll \omega_\text{av}$ so the diagonal matrix elements of  $\hat{H}_F$ are
much larger than the off-diagonal couplings. The unperturbed Floquet eigenspectra is given by $\{\pm\omega_\text{av}/2 + l\omega|l = 0, \pm 1, \pm 2, ...\}$. Degeneracies occur in the spectra when $\omega = \omega_\text{av}/k$, i.e. $T = 2\pi k /\omega_\text{av}$. Figure~\ref{Fig1}(b) shows the unperturbed eigenspectra in a scan of DD period, $T$. 

The off-diagonal terms in the $\hat{V}_{t_p} $ coupling matrix are $A_\perp f_z^k/2$ and $A_\perp f_\perp^k /2 = A_\perp(f_x^k +if_y^k)/2$ where $f_i^k$ are the Fourier amplitudes of the modulation functions. Figure~\ref{fig: pulses} shows the magnitude of the Fourier amplitudes for an XY8 pulse sequence evaluated at the degeneracies $T = 2\pi k/\omega_\text{av}$. The coherence dips expected from DD based sensing are caused by $f_z^k$, thus  for an XY8 sequence dips will occur at $T = 2\pi k / \omega_\text{av}$ for $k = 4, 12, 20,...$ as seen in Fig.~\ref{Fig1}(c). When a finite pulse is modelled the $f_\perp^k$ become non-zero opening new avoided crossings and causing new coherence dips. The resulting perturbed eigenspectra with avoided crossings were previously illustrated in Fig.~\ref{Fig1}(d).

\subsection{Calculating the coherence function}
\label{sec: Coherence} 

To model a \emph{spurious} dip $\hat{H}_F$ must be diagonalised at $T \approx 2\pi k/\omega_\text{av}$ with $k$ such that $f_\perp^k = |f_\perp^k|\exp(i\phi_\perp^k) \neq 0$. At this value of $T$ the Floquet Hamiltonian can be treated as an infinite set of $2\times 2$ matrices as the pairs of degenerate diagonal entries, $\omega_\text{av}/2 + l\omega$ and $-\omega_\text{av}/2 + (l+ k)\omega$, are coupled by the $f_\perp^k$ (or $f_\perp^{-k}$) terms whereas all other off-diagonal entries do not couple a degeneracy and can be neglected. The  procedure is shown in Fig.~\ref{fig:HamFloq} which  illustrates the specific case of a $k=2$ dip.  The Floquet Hamiltonian is effectively decoupled into $2\times 2$ subspaces and these can be diagonalised analytically. 

The Floquet Hamiltonian can then be written in the form $\hat{H}_F = \hat{D}_F\hat{\Lambda}_F\hat{D}_F^{-1}$, where $\hat{\Lambda}_F$ is the diagonal matrix of Floquet quasienergies and $\hat{D}_F$ contains all the Floquet Hamiltonian eigenstates. The stroboscopic evolution operator is given by $\hat{U}(N_pT) = \hat{D}\exp(-i\hat{\Lambda} N_pT)\hat{D}^{-1}$ where $\langle m_s\alpha|\hat{D}|m'_s\alpha'\rangle = \sum_l \langle m_s\alpha l|\hat{D}_F|m'_s\alpha' 0\rangle$ and $\langle m_s\alpha|\hat{\Lambda}|m_s\alpha\rangle = \langle m_s\alpha 0|\hat{\Lambda}_F|m_s\alpha 0\rangle$. This is valid for arbitrary $N_p$ (and given explicitly in \ref{app: coherence}) and thus an analytic expression for the spurious coherence response of the NV center can be obtained:
\begin{equation}
 \mathcal{L}(N_pT) \propto \langle \hat{S}_x \rangle = \text{Tr}[\hat{S}_x\hat{U}(N_pT)\rho_0\hat{U}^\dagger(N_pT)].
\end{equation}

Typically the NV is initially prepared in a superposition state along the $+ x$-direction and the nuclear target is assumed to be in a thermal state. The initial density matrix is $\rho_0 = \frac{1}{4}(\mathds{I} + \sigma_x)_\text{NV} \otimes \mathds{I}_\text{target}$.

The spurious coherence dip is given by
\begin{multline}
\label{eq: Lspurious}
\mathcal{L}_s(N_pT) = 1 - 2\left[\frac{\epsilon_s(k;T)^2 - (\omega_\text{av} - k\omega)^2/4}{\epsilon_s(k;T)^2}\right]   \\
\times \sin^2\left( N_p\epsilon_s(k;T)T\right) \cos^2(\phi^k_\perp + \phi_g),
\end{multline}
where $\epsilon_s(k;T) = \frac{1}{2}\sqrt{(\omega_\text{av} - k\omega)^2 + |A_\perp f_\perp^k|^2}$ describes the $k$-th avoided crossing. The new parameter $\phi_g$ is introduced to denote a global phase added to all the pulses while maintaining the phase of the initial NV state and measurement. Recent work on spurious coherence dips has shown a dependence on global phase \cite{haase2016pulse} and it is modelled here. Figure~\ref{fig: DipZooms} shows a good fit of this analytic expression to numerical data. 

At the dip $\epsilon_s(k;T^k_\text{dip}) = \frac{1}{2}|A_\perp f_\perp^k|$ where $|A_\perp f_\perp^k|$ is the width of the quasienergy avoided crossing and defines the effective coupling strength of the nuclear spin to the NV under DD control, as opposed to the bare coupling strength $A_\perp$. The term in square brackets reaches unity at the dip position so the depth of the dip is given by $\mathcal{L}_s(N_pT^k_\text{dip}) = 1 - 2\sin^2\left(N_p\frac{1}{2}|A_\perp f_\perp^k| T^k_\text{dip}\right) \cos^2(\phi^k_\perp + \phi_g)$ which reaches a maximum depth of $\mathcal{L}_s(N^{\text {max}}_pT^k_\text{dip})= -\cos( 2 (\phi^k_\perp + \phi_g))$ at $N_p = N^{\text {max}}_p = \pi/(|A_\perp f_\perp^k| T^k_\text{dip})$. The depth of spurious dips is therefore limited by the phase of $f_\perp^k$. For XY8, $\phi_\perp^k \in \{\pm \pi/4, \pm 3\pi/4\}$ so the spurious dips will never drop below zero (for $\phi_g = 0$). However, scanning $\phi_g$ gives control over the spurious dip depths. In fact the $k$'th spurious resonance can be turned off, to avoid ambiguities with other nuclear signals, by setting $\phi_g = -\phi_\perp^k \pm \pi/2$. Conversely, the contrast of a particular spurious dip can be maximised by choosing $\phi_g = -\phi_\perp^k$. Table \ref{table: mimics} lists isotopes with fundamental signals that can be mimicked by the spurious harmonic of another isotope and lists the global phase required to suppress this spurious signal. These global phases are specific for the XY8 sequence but similar relations can be derived for any other DD sequence.

\begin{table}[]
\centering
\caption{Isotopes susceptible to ambiguous characterisation due the presence of another isotope that mimics the signal at the listed harmonic. Applying the global phase, $\phi_g$, to all pulses in the XY8 sequence suppresses the unwanted spurious signal for unambiguous nuclear species classification. (For XY8 the $n\times$ harmonic of the fundamental signal is at $T^k_\text{dip} = 2\pi k/\omega_\text{av}$ for $k = 4/n$.)}
\label{table: mimics}
\renewcommand{\arraystretch}{1.5}
\begin{tabular}{c|c|c|c}
Isotope & Mimic & Harmonic  & $\phi_g$ \\ \cline{1-4}
$^1$H      & $^{13}$C                     & $4\times$   & $-\pi/4$        \\
$^{29}$Si    & $^{13}$C                   & $4/5\times$  & $-\pi/4$            \\
$^{31}$P     & $^1$H                      & $2/5\times$ & $+\pi/4$            
\end{tabular}
\end{table}

For $N_p > N^{\text {max}}_p$,  the dip simply acquires additional sideband structures, which, with increasing $N_p$, progressively fill  an envelope $\mathcal{L}^\text{env}_s(T) = 1 - 2\left[\frac{\epsilon_s(k;T)^2 - (\omega_\text{av} - k\omega)^2/4}{\epsilon_s(k;T)^2}\right]\cos^2(\phi^k_\perp + \phi_g)$ function,
which is independent of $N_p$. We also find that this coherence envelope function
 has width $W_T = 2|A_\perp f_\perp^k|T^k_\text{dip}/\omega_\text{av}$. 

A formula for the expected coherence dips can also be obtained by diagonalising the Floquet Hamiltonian at $T^k_\text{dip}$ for $k$ such that $f_z^k \neq 0$. This is given by 
\begin{equation}
\label{eq: Lnormal}
\mathcal{L}(N_pT) =  1 - 2\left[\frac{\epsilon(k;T)^2 - (\frac{\omega_\text{av} - k\omega}{2})^2}{\epsilon(k;T)^2}\right]\sin^2\left(N_p\epsilon(k;T) T\right),
\end{equation}
where $\epsilon(k;T) = \frac{1}{2}\sqrt{(\omega_\text{av} - k\omega)^2 + |A_\perp f_z^k|^2}$ describes the Floquet eigenvalue avoided crossing and there is no dependence on the phase of $f_z^k$ or a global phase. The parameters in this case have the same functional form but employ $f_z^k$ instead of $f_\perp^k$. 

Since $f_\perp^k \ll f_z^k$,  the width of the `spurious' dips is much less that those of `expected' dips. This 
implies  an increase in resolution and a possible aid for improving sensing experiments. This is described in Section \ref{sec: Resolution}.

\subsection{Comparisons with numerics}
\label{sec: Numerics}

\begin{figure}[h]
\begin{center}
\includegraphics[width=86mm]{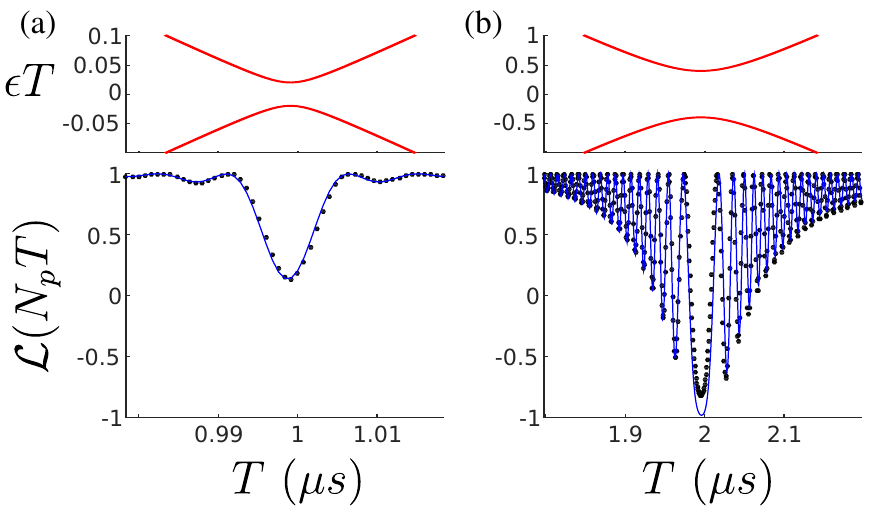}
\caption{Comparison of numerics (black dots) to the analytic expressions (blue lines) for \textbf{(a)} the $k=2$ spurious dip and \textbf{(b)} the $k = 4$ expected dip under XY8 control. Above are the associated Floquet avoided crossings. Note that the scale on the expected avoided crossing is 10 times larger. For numerical simulations here: in the Zeeman basis $\omega_z = 2\pi\times  2\text{ MHz}$ and $A_x = 2\pi\times  200\text{ kHz}$. The pulse sequence has $N_p = N/8 = 60$ repetitions. The finite pulses have height $\Omega = 2\pi\times  20\text{ MHz}$ and duration $t_p = \pi/\Omega$.}
\label{fig: DipZooms}
\end{center}
\end{figure}

To validate the new analytic expression for coherence dips it was compared against a full numeric calculation. An example of the accurate fit is shown in Fig.~\ref{fig: DipZooms}. Both, expected and spurious dips are shown to be modelled well by the analytic formula. The numerics were obtained by directly propagating the NV-nuclear spin system in the Zeeman basis. Intervals of free propagation under $\hat{H}_0$ were concatenated with the propagation during the pulse under $\hat{H}_0 + \hat{H}_p(t')$.

Notice that the the $k=2$ spurious avoided crossing is approximately $20$ times narrower than the expected $k=4$ avoided crossing. This equates to a twenty times reduction in dip width, as can be seen in the coherence traces. The spurious dips are much narrower than the expected dips. In the next section we describe a protocol to exploit this feature to obtain increased resolution in sensing experiments.

\section{Spurious dips as a resource}
\label{sec: Resolution}

Recent theory proposals and experiments make use of tunable decoupling sequences to gain control over the effective coupling to nuclear spins \cite{casanova2015robust, wang2016positioning, ma2015resolving, zhao2014dynamical}. This involves the application of sequences made of composite pulses with adjustable interpulse spacing so that one can tune $|f_z^k|$ and open or close avoided crossings, i.e. augmenting or diminishing the effective coupling to nuclear spins and thus the width of experimental signals. By choosing $|f_z^k|$ to be small one can sharpen a selected coherence dip allowing for better resolution between isolated nuclear spins. The number of pulses that can be applied is constrained by the $T_2$ coherence time so the minimal limit for $|f_z^k|$ is determined by the requirement that the dip still obtain a visible contrast after the maximum number of pulses.

One result of this work has been to see that spurious dips are naturally much narrower than the normal dips because $|f_\perp^k| \ll |f_z^k|$.  Hence, if spurious signals are observed in the spectra one can increase the spectral resolution without complex pulse sequence design. One requires only that $|f_\perp^k|$ is not so small that the dips cannot obtain appreciable contrast after the maximum allowed number of pulses is applied (this is the same for tunable pulse sequences). From another point of view, the enhanced spectral resolution by spurious dips is realized by making the effective coupling smaller than the frequency separation. Otherwise, the strong coupling invalidates the rotating wave approximation to remove perturbation from unwanted spins~\cite{casanova2015robust, wang2016positioning}.

This strategy has different benefits with respect to the use of more complex sequences. On the one hand,  the use of sequences containing composite pulses requires one to apply a number of pulses larger than the one used by a standard sequence, such as XY8. In this respect, and even under moderate pulse error conditions, the applicability of composite sequences could be challenging because the error accumulates, damaging the signal. On the other hand, the use of a robust sequence of composite pulses requires an accurate control of each pulse phase~\cite{casanova2015robust, wang2016positioning}, i.e. each pulse has to rotate the NV state around a different axis on the $xy$ plane, and this is another experimental requirement to be addressed. Note that for a XY8 sequence just two phases are needed. Furthermore, the analytic expression, Eq.~\eqref{eq: Lspurious}, indicates that one can also gain control over the spurious dips. By applying a global phase, $\phi_g$, to all pulses or by carefully designing the pulse sequence to selectively control $f_\perp^k$.

\begin{figure}[ht!]
\begin{center}
\includegraphics[width = 86mm]{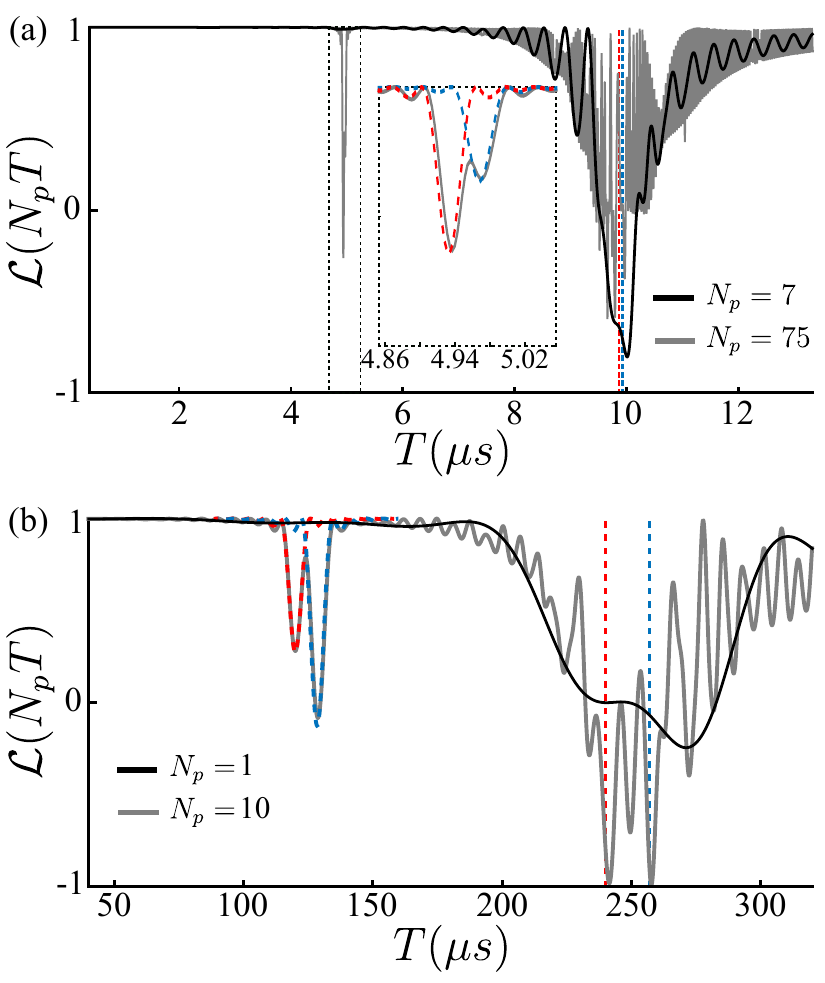}
\caption{\textbf{(a)} Numerical simulation of the NV coherence coupled to two independent nuclear spins with $\omega_\text{av}^{(1)} = 2\pi\times  402.6$ kHz and $\omega_\text{av}^{(2)} = 2\pi\times  405.4$ kHz and hyperfine coupling strengths $A_\perp^{(1)} = 2\pi\times  21.6$ kHz and $A_\perp^{(2)} = 2\pi\times  31.0$ kHz. Solid black line: The coherence trace after $N_p = 7$ repetitions of the XY8 sequence. Solid grey line: The coherence trace after $N_p = 75$ repetitions. The finite pulses have height $\Omega =2\pi\times  10$ MHz and width $t_p = \pi/\Omega$. The inset shows a magnification of the spurious dips with the analytic expression, Eq.~\eqref{eq: Lspurious},  for each dip plotted also (dashed line). 
\textbf{(b)} Numerical simulation of the NV coherence coupled to two, more remote, independent nuclear spins with $\omega_\text{av}^{(1)} = 2\pi\times  16.67$ kHz and $\omega_\text{av}^{(2)} = 2\pi\times  15.56$ kHz and hyperfine coupling strengths $A_\perp^{(1)} =2\pi\times   1.63$ kHz and $A_\perp^{(2)} = 2\pi\times  2.14$ kHz. Solid black line: The coherence trace after $N_p = 1$ repetitions of the XY8 sequence. Solid grey line: The coherence trace after $N_p = 10$ repetitions. The finite pulses have height $\Omega =2\pi\times   100$ kHz and width $t_p = \pi/\Omega$. The analytic expression, Eq.~\eqref{eq: Lspurious},  for the spurious dips is also plotted (dashed line). 
In each case the expected positions of the fundamental dip for each spin are denoted by the red and blue vertical dashed lines but these are unresolved because the frequency separation is less than the hyperfine coupling strength. By increasing the number of pulses the pairs of spins can be clearly resolved at the $k=2$ spurious dip. A global phase of $\phi_g = -\pi/4$ has been applied to the pulse sequence to enhance the contrast at the spurious dip.}
\label{fig: resolution}
\end{center}
\end{figure}

A simple protocol for increasing resolution is demonstrated in Fig.~\ref{fig: resolution}. Here an XY8 sequence drives an NV center coupled to two isolated nuclear spins ($^{13}\text{C}$ in diamond for demonstration), in the first instance the spins are $\approx 0.6$ nm from the NV whilst in the second instance the spins are more weakly coupled at $\approx 2$ nm from the NV. In both cases initial attempts to resolve the spin with the fundamental dip fail because the hyperfine coupling strengths are greater than the signal separation. However, by judiciously increasing the number of pulses and reducing the microwave pulse height (i.e. the Rabi frequency) one obtains sharp spurious features where the signals can be resolved. One can also add a global phase to the pulse sequence to enhance the contrast of the desired spurious dip.

\section{Conclusion}
\label{sec: Conclusion}

We propose the use of the microwave pulse width, $t_p$ in dynamical decoupling sequences as an additional experimental parameter in nanoscale NMR and MRI experiments. We demonstrate increased resolution in the detection of single nuclear spins by exploiting so called spurious signals, that arise due to finite pulse widths. We also propose methods for suppressing the spurious signals from nuclei when they mimic signals from other isotopes, thus removing ambiguity in nuclear spin identification.

Floquet analysis was applied as the natural framework for studying quantum sensing experiments under periodic dynamical decoupling control. It reveals a landscape of avoided crossings in an underlying quantum spectrum that control the position and depth of characteristic NV coherence dips. It was shown that adjusting $t_p$ from zero opens new avoided crossings that correspond to spurious signals in coherence traces. An analytic expression for spurious dips is given in Eq.~\eqref{eq: Lspurious} and it is shown to be in good agreement with numerics. Our study heralds a new generation of pulse sequence design that exploits realistic pulse profiles alongside tuned pulse spacings and phases.

\section*{ACKNOWLEDGEMENTS}

The authors acknowledge valuable discussions with Renbao Liu. J.L. acknowledges an EPSRC DTA studentship. J.C. acknowledges support to the Alexander von Humboldt foundation. This work was supported by the ERC Synergy grant BioQ, the EU projects DIADEMS and EQUAM and the DFG Collaborative research center TRR 21.

\appendix

\section{Average Hamiltonian states for single spin sensing}
\label{app: AvHam}

\begin{figure*}
$$
\hat{H}'(t) = 
\begin{pmatrix}
\frac{\omega_\text{av}}{2} & \frac{A_\perp}{2} f_z(t) & 0 & \frac{A_\perp}{2}[f_x(t) -i f_y(t)]\\ 
\frac{A_\perp}{2} f_z(t) & -\frac{\omega_\text{av}}{2} & \frac{A_\perp}{2}[f_x(t) -i f_y(t)] & 0\\ 
0 & \frac{A_\perp}{2}[f_x(t) +i f_y(t)] & \frac{\omega_\text{av}}{2} & -\frac{A_\perp}{2} f_z(t)\\ 
\frac{A_\perp}{2}[f_x(t) +i f_y(t)] & 0 & -\frac{A_\perp}{2} f_z(t) & -\frac{\omega_\text{av}}{2}
\end{pmatrix}
$$
\caption{Full NV - nuclear spin Hamiltonian in the average nuclear spin basis. It is composed of the diagonal average Hamiltonian with periodic off-diagonal terms.}
\label{fig: IntHam}
\end{figure*}

To analyse the effect of finite pulses we model an experiment sensing a single nuclear spin half, $\hat{\textbf{I}}$. For nuclear spins far from the NV center the hyperfine coupling can assumed to be pure dephasing. In this case the nuclear spin Hamiltonian is conditioned on the state of the NV center: $\hat{H}^{(0)} = \gamma_n B_0\hat{I}_{z_{\text{NV}}}$ and $\hat{H}^{(1)} = \gamma_n B_0\hat{I}_{z_{\text{NV}}} + \textbf{A}\cdot \hat{\textbf{I}}$ where the superscripts 0 and 1 denote the NV center state. Here $\hat{\textbf{z}}_\text{NV}$ is set by the NV axis and $\hat{\textbf{x}}_\text{NV}$ is set by the hyperfine field such that $\textbf{A} = (A_x, 0, A_z)$. A magnetic field, $\textbf{B} = B_0 \hat{\textbf{z}}_\text{NV}$, is applied parallel to the NV axis and $\gamma_n$ is the nuclear gyromagnetic ratio. In the lab frame the nuclear average Hamiltonian is given by $\hat{H}_\text{av} = \gamma_n B_0\hat{I}_{z_{\text{NV}}} + \frac{1}{2}\textbf{A}\cdot \hat{\textbf{I}}$ and the nuclear interaction Hamiltonian is $\hat{V} = \frac{1}{2}\textbf{A}\cdot \hat{\textbf{I}}$.

In the average Hamiltonian basis $\hat{H}_\text{av} = \omega_\text{av}\hat{I}_z$ and $\hat{V} = A_\perp\hat{I}_x + A_\parallel\hat{I}_z$ where $\hat{\textbf{z}} = \cos\theta_\text{av}\hat{\textbf{z}}_\text{NV} + \sin\theta_\text{av}\hat{\textbf{x}}_\text{NV}$ and $\hat{\textbf{x}} = \cos\theta_\text{av}\hat{\textbf{x}}_\text{NV} - \sin\theta_\text{av}\hat{\textbf{z}}_\text{NV}$. The average Hamiltonian frequency is $\omega_\text{av}= \sqrt{(\gamma_n B_z + A_z/2)^2 + (A_x/2)^2}$ and the hyperfine components in this basis are given by $A_\perp = (A_x\cos\theta_\text{av} - A_z\sin\theta_\text{av})/2$ and $A_\parallel = (A_z\cos\theta_\text{av} + A_x\sin\theta_\text{av})/2$ where $\theta_\text{av} = \arctan(A_x/(2\gamma_nB_0 + A_z))$. Figure~\ref{fig: physicalpic} shows the system fields in the two bases. For simplicity we ignore the $A_\parallel$ contribution to the interaction term as for weak coupling the effect of parallel fluctuations is small. In the Floquet Hamiltonian this can be understood as it does not couple degenerate energy levels. Perturbative techniques could be used to study the effect of parallel coupling strengths.

Considering the application of a repeated XY8 dynamical decoupling sequence the NV - nuclear Hamiltonian can be written in the frame rotating under the effect of the pulse Hamiltonian as
\begin{equation}
\label{eq: Hamiltonian}
\hat{H}'(t) = 
\hat{\mathds{I}} \otimes \omega_\text{av}\hat{I}_z + \sum_{i = x,y,z}f_i(t)\hat{\sigma}_i \otimes A_\perp\hat{I}_x,
\end{equation}
where the modulation functions for XY8 are shown in Fig.~\ref{fig: pulses}. This Hamiltonian is shown in matrix form in Fig.~\ref{fig: IntHam} so that it can be compared with a schematic representation of $\hat{H}_F$, shown in Fig.~\ref{fig:HamFloq}. The elements of the Floquet Hamiltonian, $\hat{H}_F$, are obtained from Eq.~\eqref{eq: FloquetHam}. In the weak coupling regime, $\omega_\text{av} \gg A_\perp$, $\hat{H}_F$ is approximately diagonal with only small off-diagonal perturbations.

The unperturbed eigenvalues are given by the diagonal entries $\epsilon_{\pm l} = \pm\omega_\text{av}/2 + l\omega$, which are doublets. These doublets remain degenerate as there is no off-diagonal perturbation connecting them. More interesting are the degeneracies that occur when the DD frequency, $\omega$, is set to $\omega = \omega_\text{av}/k$. At this point $\omega_\text{av}/2 + m\omega = -\omega_\text{av}/2 + (m + k)\omega$ for all $m$. In Fig.~\ref{Fig1}(b) these unperturbed eigenvalues are plotted as $\epsilon_{\pm l}T = \pm\omega_\text{av}T/2 + l2\pi$. 

If the off-diagonal perturbation connecting two diagonal entries is non-zero the degeneracy will be lifted and an avoided crossing will appear. These perturbed eigenvalues are plotted in Fig.~\ref{Fig1}(c),(d).

\section{Coherence functions}
\label{app: coherence}

\subsection{$t_p=0$ case: `Expected' coherence dips} 

The Fourier amplitudes, $f_z^k$ determine the positions of the expected coherence dips. The Fourier series for an XY8 sequence are shown in Fig.~\ref{fig: pulses}(c). Dips appear when $\omega = \omega_\text{av}/k$ for $k = 4,12,20,..$. At these values of $\omega$ we must take the off-diagonal perturbations into account and the Floquet Hamiltonian separates into an infinite set of $2\times 2$ matrices which can be diagonalised.

As discussed in Section \ref{sec: Coherence} the stroboscopic evolution in the original Hilbert space is obtained from the diagonalisation of $\hat{H}_F$. At the expected coherence dips this is given by 
\begin{equation}
\hat{U}(N_pT) = e^{-ikN_p\pi}
\begin{pmatrix}
u_a &u_b &0 &0 \\
-u_b^* &u_a^* &0 &0 \\
0 &0 &u_a & -u_b \\
0 &0 & u_b^* & u_a^*
\end{pmatrix},
\end{equation}
where $u_a = \cos( N_p\epsilon(k;T)T) - i\sin( N_p\epsilon(k;T) T) \cos\theta_F(k;T)$,  $u_b = -i \sin( N_p\epsilon(k;T) T) \sin\theta_F(k;T)$ and $\epsilon(k;T) = \frac{1}{2}\sqrt{(\omega_\text{av} - k\omega)^2 + |A_\perp f_z^k|^2}$ and $\theta_F(k;T) = \arctan(|A_\perp f_z^k|/(\omega_\text{av} - k\omega))$.

The coherence of the NV sensor near to the dip at $T^k_\text{dip} = 2k\pi/\omega_\text{av}$ is thus modelled by
\begin{equation}
\mathcal{L}(N_pT) = 1 - 2\sin^2(N_p\epsilon(k;T) T) \sin^2 \theta_F(k;T).
\end{equation}

In the main text we use the identity $\sin^2 \theta_F(k;T) = [\epsilon(k;T)^2 - (\omega_\text{av} - k\omega)^2/4]/\epsilon(k;T)^2$ to emphasise the importance of the Floquet quasienergy spectra, $\epsilon(k;T)$.

\subsection{$t_p\neq 0$ case: `Spurious' coherence dips}

For the XY8 sequence there are spurious dips at $T^k_\text{dip} = 2k\pi/\omega_\text{av}$ for $k = 1,2,3,5,6,7,9,10,11,...$ as shown in Fig.~\ref{Fig1}. For these $k$ the Floquet Hamiltonian will again have degeneracies in diagonal terms lifted by off-diagonal perturbations. The Floquet Hamiltonian seperates into $2 \times 2$ matrices and can be diagonalised. The stroboscopic evolution is then given by
\begin{equation}
\hat{U}_s(N_pT) = e^{-ikN_p\pi} 
\begin{pmatrix}
v_a &0 &0 &v_b \\
0 &v_a^{*} &-v_b &0 \\
0 &v_b^{*} &v_a & 0 \\
-v_b^{*} &0 & 0 & v_a^{*}
\end{pmatrix},
\end{equation}
where $v_a = \cos( N_p\epsilon(k;T) T )- i\sin( N_p\epsilon_s(k;T) T )\cos\theta_{Fs}(k;T)$, $v_b=-i \sin (N_p\epsilon_s(k;T) T) \sin\theta_{Fs}(k;T) e^{-i\phi_F^k}$ and $\epsilon_s(k;T) = \frac{1}{2}\sqrt{(\omega_\text{av} - k\omega)^2 + |A_\perp f_\perp^k|^2}$,  $\theta_{Fs}(k;T) = \arctan(|A_\perp f_\perp^k|/(\omega_\text{av} - k\omega))$ and $\phi_F^k = \phi^k_\perp$ where $f_\perp^k = |f_\perp^k|e^{i\phi^k_\perp}$.

The spurious dip near to $T^k_\text{dip} = 2k\pi/\omega_\text{av}$ in an NV coherence trace is thus modelled by
\begin{equation}
\mathcal{L}_s(N_pT) = 1 - 2\sin^2(N_p\epsilon_s(k;T) T) \sin^2 \theta_{Fs}(k;T)\cos^2\phi_F^k.
\end{equation}

\section{CPMG sequences}
\label{app: CPMG}

\begin{figure}[ht!]
\begin{center}
\includegraphics[width = 82mm]{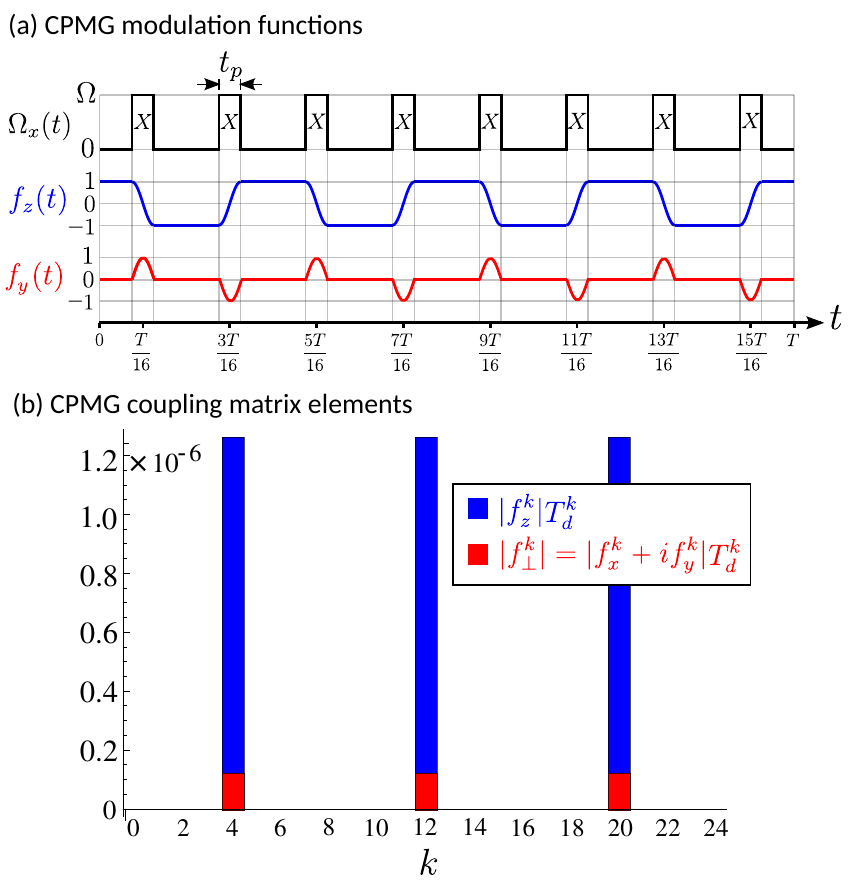}
\caption{{\bf (a)} Shows the modulation functions for a CPMG sequence with pulses of finite duration $t_p$.{\bf (b)} Matrix elements of the modulations function $\langle l+k|f_{i}|l\rangle$ in the dressed state basis. For CPMG the spurious couplings $f_\perp^k$ coincide with the expected couplings $f_z^k$. Thus the presence of finite pulses only perturbs slightly the expected signal and does not produce new signals at spurious harmonics (as for XY8).} 
\label{fig: CPMG}
\end{center}
\end{figure}

In this section we briefly analyse the sensor response under the Carr-Purcell-Meiboom-Gill (CPMG) dynamical decoupling sequence. A CPMG sequence has the same pulse positions as XY8 except all pulses have the same (x) phase. The pulse amplitudes are shown in Fig.~\ref{fig: CPMG}(a) along with the modulation functions obtained from Eq.~\eqref{eq: modfuncs}. The $f_z(t)$ modulation function is independent of the pulse phases thus is the same as for XY8. The change of pulse phase is encoded into the perpendicular modulation functions. For CPMG $f_x(t) = 0$ whilst $f_y(t)$ is shown.

When detecting a single nuclear spin, as in the main text, one first studies the unperturbed Floquet spectrum to find there are degeneracies at $T = 2\pi k/\omega_\text{av}$ (We choose an 8 pulse CPMG unit as one period to allow a simple comparison with XY8). Upon calculating the coupling matrix elements in the Floquet Hamiltonian one finds that the spurious couplings $A_\perp f_\perp^k/2$ are non-zero for $k = 4, 12, 20, ...$ coinciding with the non-zero expected couplings $A_\perp f_\perp^k/2$, see Fig.~\ref{fig: CPMG}(b). This is due to $f_z(t)$ and $f_y(t)$ having the same frequency. This immediately explains why spurious resonances are not seen in CPMG coherence traces. The spurious signal is buried inside the expected dip. In fact the finite pulse effect in CPMG traces is to perturb the expected coherence dip but not create new signals.

The coherence can again be modelled by $\mathcal{L}(N_pT) \propto \text{Tr}[\hat{S}_x\hat{U}(N_pT)\rho_0\hat{U}^\dagger(N_pT)]$ where the propagator at time $t = N_pT$ is determined by an effective Hamiltonian $\hat{U}(N_pT) = \exp(-i\hat{H}_\text{eff}N_pT)$. At $T = 2\pi k/\omega_\text{av}$:
\begin{equation}
\hat{H}_\text{eff} = 
\begin{pmatrix}
\frac{\omega_\text{av} - k\omega}{2} &\frac{A_\perp }{2}f_z^k &0 &\frac{A_\perp}{2}f_\perp^{k*} \\
\frac{A_\perp }{2}f_z^k &-\frac{\omega_\text{av} - k\omega}{2} &\frac{A_\perp}{2}f_\perp^{-k*} &0 \\
0 &\frac{A_\perp}{2}f_\perp^{-k} &\frac{\omega_\text{av} - k\omega}{2} & -\frac{A_\perp }{2}f_z^k \\
\frac{A_\perp}{2}f_\perp^k &0 & -\frac{A_\perp }{2}f_z^k & -\frac{\omega_\text{av} - k\omega}{2}
\end{pmatrix}.
\end{equation}

\end{document}